\documentclass[showpacs,10pt,twocolumn,prb]{revtex4-1}

\usepackage{amsmath}
\usepackage{amssymb}
\usepackage{graphics}
\usepackage{epsfig}
\usepackage{CJK}
\usepackage{color}

\begin{document}


\title{Superconductivity in Ir$_{1-x}$Rh$_{x}$Te$_{2}$ (0 $\leqslant$ x $\leqslant $ 0.3)}
\author{Hechang Lei,$^{1,\ast }$ Kefeng Wang,$^{1}$ Milinda Abeykoon,$^{1}$ Emil Bozin,$^{1}$ J. B. Warren,$^{2}$ and C. Petrovic$^{1}$}
\affiliation{$^{1}$Condensed Matter Physics and Materials Science Department, Brookhaven
National Laboratory, Upton, New York 11973, USA\\
$^{2}$Instrumentation Division, Brookhaven National Laboratory, Upton, New
York 11973, USA}
\date{\today}

\begin{abstract}
We report superconductivity, physical and structural properties of Ir$_{1-x}$Rh$_{x}$Te$_{2}$ (0 $\leqslant $ x $\leqslant $ 0.3). Superconducting transition with maximum $T_{c}\sim$ 2.6 K appears when the doping content of Rh $x$ is in between 0.15 and 0.3. Further increasing the content of Rh will suppress the superconductivity. On the other hand, the high-temperature structural transition is suppressed gradually as Rh is incorporated into the lattice, eventually vanishing at $x=0.2$. These results imply competing relationship between structural transition and superconductivity. The isovalent ionic substitution of Rh into Ir has different effects on physical properties when compared to the isovalent anionic substitution of Se into Te, in which the structural transition is enhanced with Se substitution. It indicates the changes of structural parameters such as ionic size and electronegativity of elements will also have important effects on the evolution of physical properties in IrTe$_{2}$.
\end{abstract}

\pacs{74.62.Dh, 74.10.+v, 74.25.-q}

\maketitle

\section{Introduction}
Layered transition metal dichalcogenides (TMDCs) have been studied for a long time, in part due to competing orders such as charge density wave (CDW) state and superconductivity (SC).\cite{Wilson1}$^{-}$\cite{Morosan} The interplay between CDW and SC is the fundamental question in these systems. The CDW state can be tuned into SC via intercalation, substitution or pressure.\cite{Morosan,Liu,Sipos} Usually, there is a dome-like phase diagram, i.e., the CDW transition temperature decreases when the superconducting critical temperature $T_{c}$ increases. It indicates that CDW and SC compete.\cite{Morosan,Liu,Sipos} On the other hand, there are other TMDCs in which CDW and SC coexist.\cite{Gabovich} Therefore, the relationship of these two kinds of quantum orders is not simple and is material and crystal structure dependent.

Charge order and superconductivity are also important ingredient in the physics of high $T_{c}$ cuprate oxides, yet both are much more complicated and are found in proximity to strong magnetic interactions.\cite{Daou, Tacon,Ghiringhelli,Torchinsky,Chang} Since layered TMDCs offer less complex crystal structure in the absence of magnetic order, they can be used to study CDW and SC at a simpler stage.

The discovery of superconductivity in Pt, Pd, Cu substituted/intercalated CdI$_{2}$-type IrTe$_{2}$ with $T_{c}$ up to about 3 K has triggered a renewed interest in this field.\cite{Pyon,Yang JJ,Kamitani} IrTe$_{2}$ shows a structural transition from a trigonal to a monoclinic phase when cooled across $\sim$ 250 K.\cite{Matsumoto} The transmission electron microscope (TEM), photoemission and tight-binding electronic structure calculation results show that there is the superlattice modulation with a propagation vector of $q=$ (1/5, 0, -1/5), ascribed to an orbitally driven Peierls instability.\cite{Yang JJ,Ootsuki2} However, the results from NMR, angle-resolved photoemission spectroscopy (ARPES) and optical conductivity spectra measurements suggest that this structural transition may not be due to the CDW transition but due to the reduction of the kinetic energy of Te $p$ bands.\cite{Fang AF,Mizuno,Ootsuki} On the other hand, with Pt, Pd, Cu substitution or intercalation, the high temperature structural transition is suppressed quickly and superconductivity appears at low temperature, indicating the competing relation between these two phenomena.

Isovalent substitution is an effective way to clarify the origin of the structural transition and superconductivity. It is similar to pressure because it should not induce the extra carrier density but should change the structural parameters, ionic size and electronegativity of elements, thus affecting physical properties. Previous studies have shown that the structural transition at high temperatures is enhanced while the superconducting transition is suppressed by either hydrostatic (Ir$_{1-x}$Pt$_{x}$Te$_{2}$) or chemical (IrTe$_{2-x}$Se$_{x}$) pressure.\cite{Kiswandhi,Oh} The latter results are ascribed to the stabilization of polymeric Te-Te bonds with replacing of Te with the more electronegative Se. This is different from other TMDCs exhibiting CDW/SC states where pressure usually suppresses the CDW state and enhances the superconducting state.\cite{Liu, Sipos}

In this work, we report the physical properties of CdI$_{2}$-type Ir$_{1-x}$Rh$_{x}$Te$_{2}$ ($0\leqslant x\leqslant 0.2$) polycrystalls. Our results indicate that the high-temperature structural transition is suppressed by Rh substitution gradually and the superconductivity appears at low temperature with maximum $T_{c}$ $\sim$ 2.6 K, similar to the results of electronic doping or intercalating samples. It implies that the structural parameters such as ionic size and electronegativity of elements might be important for the evolution of physical properties in Ir$_{1-x}$Rh$_{x}$Te$_{2}$.

\section{Experiment}

Polycrystalline samples of Ir$_{1-x}$Rh$_{x}$Te$_{2}$ were synthesized using a solid-state reaction method as described previously.\cite{Yang JJ} Stoichiometric amounts of Ir, Rh, and Te elements were mixed, ground, and pelletized. Then, the pellets were placed in alumina crucible which is sealed into quartz tubes with refilled 0.2 atm Argon gas. The pellets were sintered at 1000 ${{}^{\circ }}C$ for 15 h, followed by furnace cooling to room temperature. The process was repeated once with an intermediate grinding. The structure of the samples was characterized by powder X-ray diffraction (XRD) using capillary transmission geometry in 1 mm diameter cylindrical Kapton capillaries at the X7B beamline of the National Synchrotron Light Source (NSLS) at the Brookhaven National Laboratory. Samples were measured using a 0.5 mm$^{2}$ monochromatic X-ray beam of $\sim$ 38 keV (0.3916 ${\AA}$) at 300 K. A Perkin Elmer 2D detector was mounted orthogonal to the beam path 376.4 mm away from the sample. The data were collected up to Q = 4$\pi$sin$\theta$/$\lambda$ = 12 ${\AA}$$^{-1}$. The average stoichiometry was determined by energy-dispersive X-ray spectroscopy (EDX) in a JEOL JSM-6500 scanning electron microscope. Electrical transport, heat capacity and magnetization measurements were carried out in Quantum Design PPMS-9 and MPMS-XL5. Thermal transport was measured using one-heater-two-thermometer method in PPMS-9. The relative error in our measurement was $\frac{\Delta \kappa}{\kappa}\sim$5$\%$ and $\frac{\Delta S}{S}\sim$5$\%$ based on Ni standard measured under identical conditions. Sample dimensions were measured by an optical microscope Nikon SMZ-800 with 10 $\mu $m resolution.

\section{Results and Discussions}

\begin{figure}[tbp]
\centerline{\includegraphics[scale=0.35]{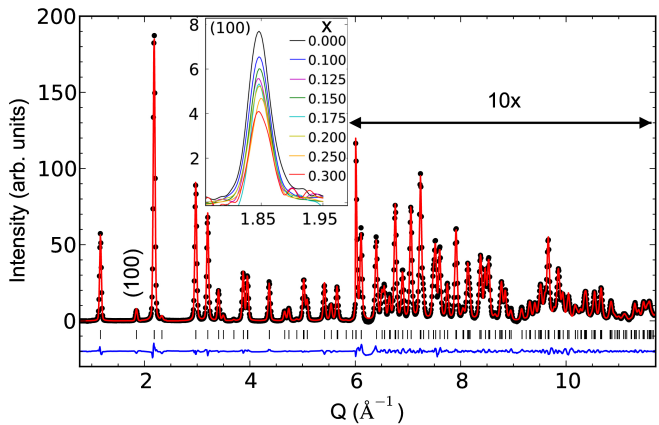}} \vspace*{-0.3cm}
\caption{The Rietveld refinement of the background subtracted IrTe$_{2}$ synchrotron powder x-ray diffraction data up to Q $\sim$ 12 ${\AA}^{-1}$. Plots show the observed (dots) and calculated (solid red line) powder patterns with a difference curve (shown at the bottom of the diagram). Vertical tick marks represent Bragg reflections in the P-3m1 space group. The inset shows evolution of the normalized intensity of (100) Bragg reflection with increasing $x$ in Ir$_{1-x}$Rh$_{x}$Te$_{2}$ for $0\leq x\leq 0.3$.
}
\end{figure}

\begin{figure}[tbp]
\centerline{\includegraphics[scale=0.50]{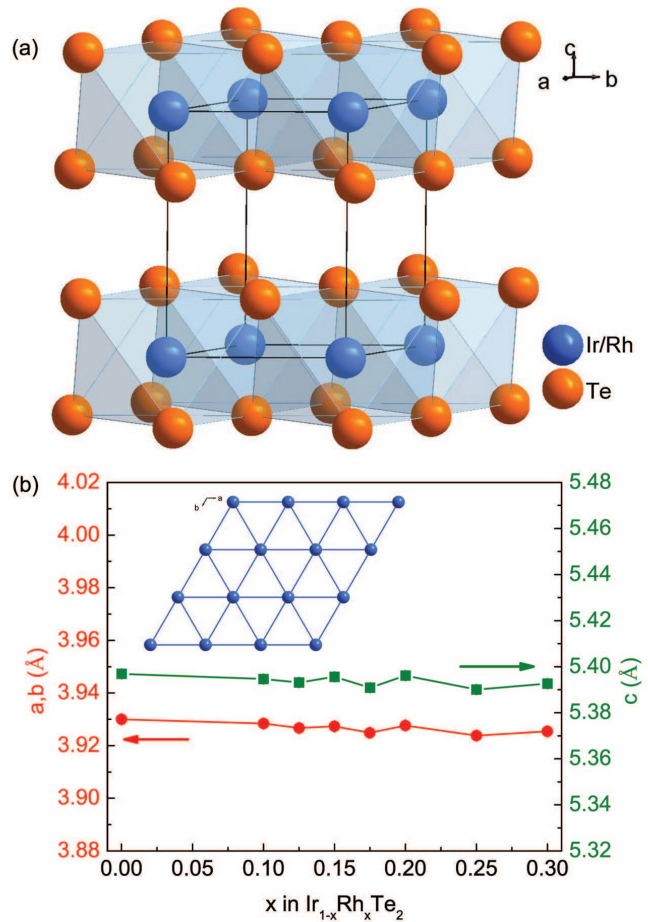}} \vspace*{-0.3cm}
\caption{(a) Crystal structure of Ir$_{1-x}$Rh$_{x}$Te$_{2}$ with Ir/Rh sites marked in blue and Te sites marked in orange. (b) Unit cell parameters as a function of Rh substitution up to $x=0.3$. Inset: the schemes of the Ir/Rh triangular lattice of the trigonal phase.}
\end{figure}

EDX stoichiometry confirmed Ir$_{1-x}$Rh$_{x}$Te$_{2}$ stoichiometry and nominal Ir/Rh ratio within up to 3$\%$ experimental error. Rietveld analysis was carried out on data sets obtained from 2D XRD images by integration into 1D patterns using Fit2d computer package.\cite{Hammersley} The refinement was performed using the General Structure Analysis System (GSAS/EXPGUI) computer package.\cite{Larson,Toby} A pseudo-Voigt function and a shifted Chebyshev polynomial were used to refine the peak profile and the background. After refining the zero-shift, lattice parameters and the background, Gaussian and Lorentzian parameters of the profile, GU, GV, GW, LX, and LY, were refined.\cite{Young} Then the atomic coordinates, occupation numbers, and the isotropic thermal displacement parameters (U$_{iso}$'s) were refined. All profile and structural parameters were refined simultaneously to optimize the quality of fits and structural models at the end of refinement. We used a room temperature CdI$_{2}$ prototype structure and trigonal symmetry (P-3m1, 1-T phase).\cite{Hockings} Figure 1 shows fits to the data with no impurity peaks present. Rietveld analysis produced excellent fits to the data up to a high Q, suggesting both high purity of samples and high quality of the XRD data.

Ir$_{1-x}$Rh$_{x}$Te$_{2}$ has a layered structure (Fig. 2(a)). There are a large number of compounds belonging to this family, especially TMDCs such as TX$_{2}$ (T = Ti, Ta, or Nb, X = S, Se, or Te). In this structure, the edge-sharing Ir/Rh-Te octahedra form Ir/Rh-Te layers in the $ab$ plane, which lead to the equilateral triangle network of Ir ions (inset in Fig. 2(b)). The Ir/Rh-Te layers stack along the $c$ axis with Te-Te bonds instead of weak van der Waals gap which has been often observed in TMDCs.\cite{Yang JJ, Lee1, Pettenkofer} Although the interlayer interaction might be stronger than in TMDCs, there are still some ions can be intercalated between Ir/Rh-Te layers, such as Pd and Cu.\cite{Yang JJ, Kamitani} The intercalation usually results in the increase of the $c$-axial lattice parameter.\cite{Yang JJ, Pyon} On the other hand, for Pt or Pd substitution where the $c$ axis decreases with doping. In contrast, the $a$ and $c$ axial lattice parameters of Ir$_{1-x}$Rh$_{x}$Te$_{2}$ series are almost unchanged with Rh substitution (Fig. 2(b)), which may be partially due to the similar ionic radius between Ir and Rh.

\begin{figure}[tbp]
\centerline{\includegraphics[scale=0.37]{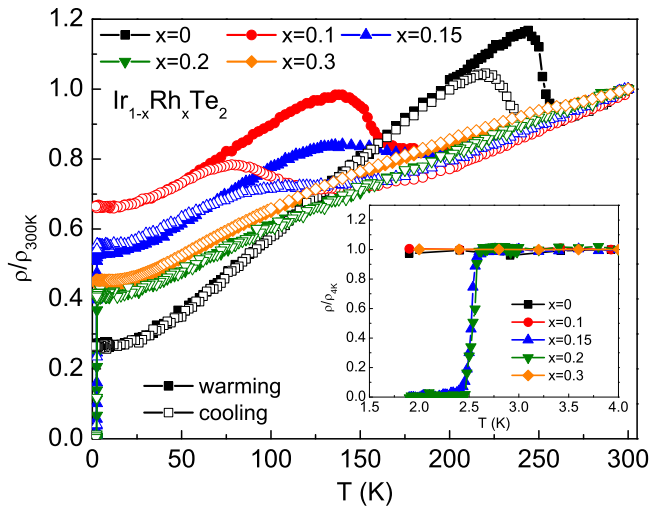}} \vspace*{-0.3cm}
\caption{Temperature dependence of the resistivity of the Ir$_{1-x}$Rh$_{x}$Te$_{2}$ (normalized at 300 K). Closed and open symbols indicate the warming and cooling process, respectively. Inset: enlarged part of temperature dependence of resistivity below 4 K (normalized at 4 K) for warming process.}
\end{figure}

As shown in Fig. 3, the temperature dependence of resistivity of pure IrTe$_{2}$ shows metallic behavior with a significant thermal hysteresis at about 250 K, which has been ascribed to the structural transition from the trigonal (P-3m1) to monoclinic symmetry (C2/m).\cite{Matsumoto} However, the origin of this structural transition is still disputed. Electron diffraction (ED) suggest that the structural transition is driven by charge-orbital density wave (DW) state with wave vector of $q=\{1/5, 0, -1/5\}$.\cite{Yang JJ} On the other hand, the NMR experiment does not provide the evidence for charge DW order and the optical spectroscopic as well as angle-resolved photoemission spectroscopy (ARPES) measurements also do not observe the gap that would correspond to the DW state near the Fermi level.\cite{Mizuno, Fang AF, Ootsuki} In addition, the theoretical calculation suggests that the structural transition is mainly caused by the evolution of Te $p$ bands rather than the instability of Ir $d$ bands, which results in a reduction of the kinetic energy of the electronic system.\cite{Fang AF, Kamitani}

With Rh substitution, the hysteresis becomes broad and shifts to lower temperature (Fig. 3). The high-temperature anomaly disappears at $x=0.2$ which is much larger substitution content when compared to Pt, Pd, or Cu substitution/intercalation where only several percent ($<$ 5\%) will suppress the structural transition completely.\cite{Yang JJ, Pyon, Kamitani} This could be due to the same valence of Rh to Ir but the origin of this phenomenon needs to be studied further. On the other hand, when $x\geq0.15$, the superconductivity emerges with the transition temperature $T_{c,onset}$ = 2.6 K. Interestingly, the transition temperature does not change much with the doping level for $0.15\leq x \leq0.2$. The transition temperature is comparable to those with other dopants or intercalating agents.\cite{Yang JJ, Pyon, Kamitani} With further increasing the content of Rh ($x=0.3$), the superconducting transition disappears completely, the dome-like dependence of $T_{c}$ on $x$ is similar to previously reported Pd, Pt and Cu substituted/intercalated IrTe$_{2}$ systems.\cite{Yang JJ, Pyon, Kamitani}

\begin{figure}[tbp]
\centerline{\includegraphics[scale=0.28]{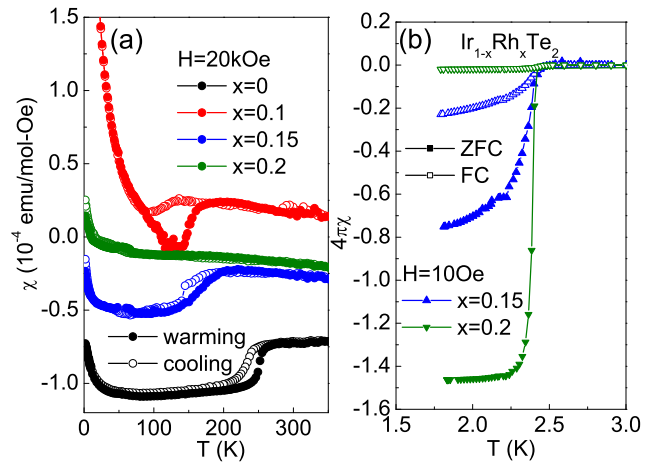}} \vspace*{-0.3cm}
\caption{(a) dc magnetic susceptibility of Ir$_{1-x}$Rh$_{x}$Te$_{2}$ samples between 1.8 and 350 K at H = 20 kOe. Closed and open symbols indicate the warming and cooling processes, respectively. The data at 350 K is subtracted for clarity ($\Delta \chi=\chi(T)-\chi(350K)$). (b) Enlarged temperature dependence of $4\pi\chi(T)$ of Ir$_{1-x}$Rh$_{x}$Te$_{2}$ for $x=0.15$ and 0.2 at $H=$ 10 Oe in zero-field-cooling (ZFC) and field-cooling (FC) modes at low temperature.}
\end{figure}

As shown in Fig. 4(a), IrTe$_{2}$ shows an anomalous magnetic susceptibility $\chi=M/H$ drop accompanied with a large thermal hysteresis between 200 - 260 K. This is consistent with the hump appearing in $\rho(T)$ curve. With Rh substitution, the $\chi(T)$ anomaly shifts to lower temperature and the magnitude of the magnetic susceptibility drop becomes weaker. When the doping level of Rh is up to 20 \%, the high temperature anomaly is completely suppressed below 1.8 K (Fig. 4(a)). Superconductivity of Ir$_{1-x}$Rh$_{x}$Te$_{2}$ for $0.1<x<0.3$ is confirmed by the magnetization measurement (Fig. 4(b)). For both $x=0.15$ and 0.2 samples, the superconducting transition temperatures in $\chi(T)$ curves are about 2.5 K, consistent with the $T_{c,onset}$ derived from $\rho(T)$ curves. The large superconducting volume fractions for both $x=0.15$ and 0.2 samples confirm the bulk superconductivity of these samples. The value of $4\pi\chi$ for $x=0.2$ sample is larger than 100 \% due to the effect of demagnetization factor of rectangular sample. On the other hand, the field-cooling (FC) magnetization of all superconducting samples are very small, indicating the strong vortex pinning in Ir$_{1-x}$Rh$_{x}$Te$_{2}$.

\begin{figure}[tbp]
\centerline{\includegraphics[scale=0.33]{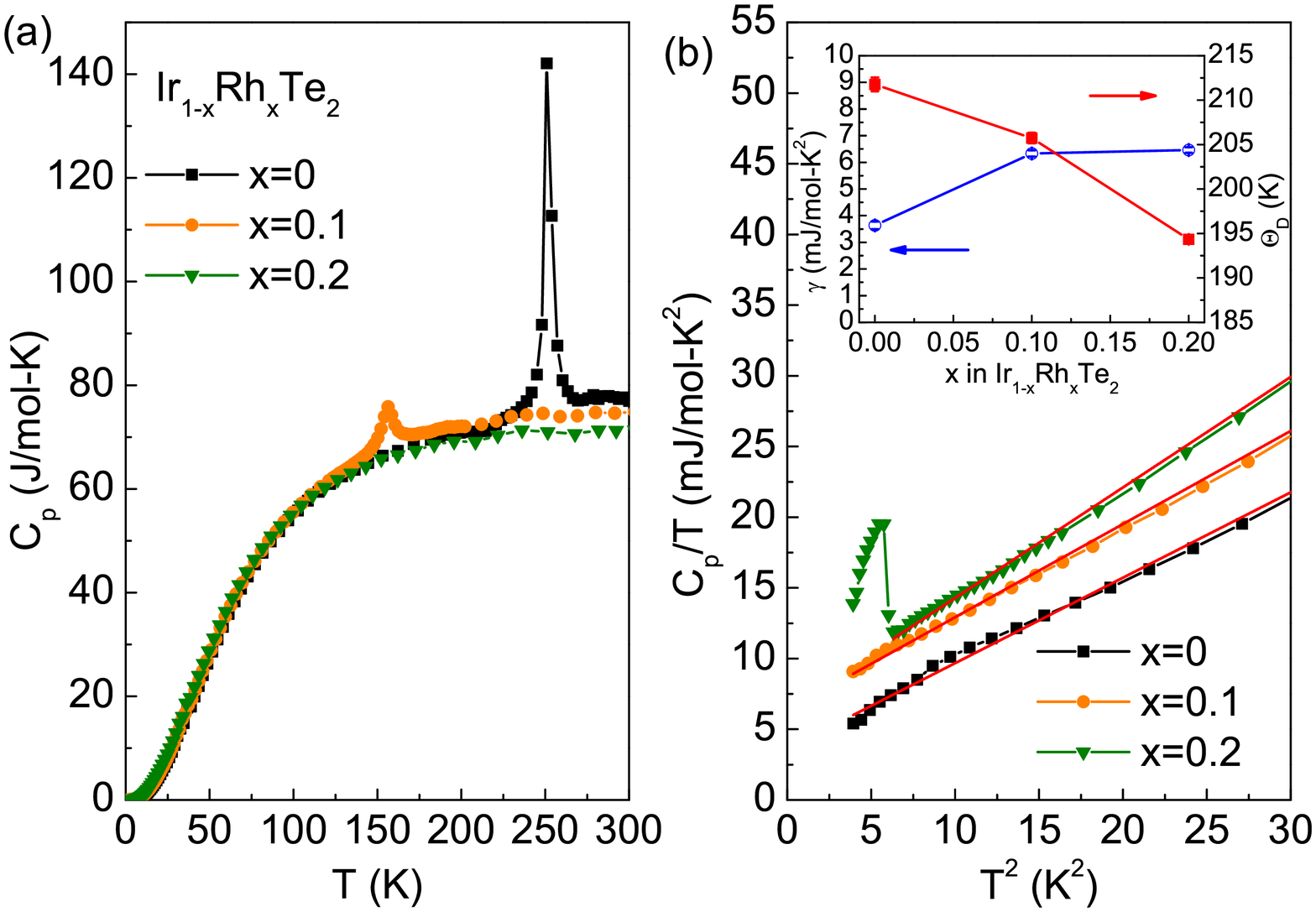}} \vspace*{-0.3cm}
\caption{(a) Temperature dependence of specific heat $C_{p}$ for Ir$_{1-x}$Rh$_{x}$Te$_{2}$. (b) Specific heat divided by temperature $C_{p}/T$ as a function of $T^{2}$ in zero field. The solid curve represents the fittings using the formula $C_{p}/T=\gamma+\beta T^{2}$. Inset: the evolution of Debye temperatures $\Theta_{D}$ and electronic specific heat coefficience $\gamma$ with Rh substitution.}
\end{figure}

Fig. 5(a) shows the specific heat of Ir$_{1-x}$Rh$_{x}$Te$_{2}$ between 1.95 and 300 K. For pure IrTe$_{2}$, there is a peak at $T\sim$ 251 K, corresponding to the anomaly in resistivity and magnetization curves. It indicates that this phase transition is first-order, consistent with the reported value in the literature.\cite{Fang AF} With Rh doping, the peak shifts to $T\sim$ 156 K and the intensity of peak also becomes weaker and less sharp than that in pure IrTe$_{2}$. It confirms that Rh doping suppresses structural transition to lower temperature and hints at possible change from first- to the second-order nature.

At low temperature (above $T_{c}$), the specific heat can be fitted very well by using the formula $C_{p}/T=\gamma+\beta T^{2}$ (red solid lines in Fig. 5(b)). The obtained $\gamma$ and derived Debye temperature $\Theta_{D}$ from $\beta$ using the relation $\Theta_{D}=(12\pi^{4}NR/5\beta)^{1/3}$, where $N=$ 3 is the number of atoms per formula unit and $R$ is the gas constant, are plotted in the inset of Fig. 5(b). The electronic specific heat of IrTe$_{2}$ is $\sim$ 3.63 mJ/mol-K$^{2}$, close to previous results.\cite{Pyon, Fang AF} When Rh is doped, the value of $\gamma$ increases to $\sim$ 6.34 mJ/mol-K$^{2}$ at $x=$ 0.1. Because the $\gamma$ is proportional to the electronic density of states (DOS) near the Fermi level, similar to Pt doping, Rh doping likely increases the area of Fermi surface companying with the suppression of structural phase transition at high temperature. It should be noted that in pure IrTe$_{2}$, this phase transition might not be driven by the orbital-driven Peierls transition.\cite{Yang JJ, Fang AF, Ootsuki} Thus, the decrease of the area of Fermi surface in IrTe$_{2}$ at low temperature might not be due to the gapping of Fermi surface originating from DW transition of Ir atoms, but because of the reconstruction of Fermi surface caused by the crystal field effect on Te atoms.\cite{Fang AF} With further increase in Rh content, the value of $\gamma$ increases slightly to 6.46 mJ/mol-K$^{2}$ at $x=$ 0.2, which is different from the Pt doping where $\gamma$ starts to decrease when $x\geq$ 0.04. It is ascribed to the decrease of DOS of IrTe$_{2}$ above the Fermi level and the shift the Fermi level upward due to the partial substitution of Pt for Ir.\cite{Pyon} In contrast, assuming that the rigid band model is valid, because Rh is isovalent to Ir, the substitution of Rh into Ir could have only minor effects on the Fermi level, resulting in smaller changes of DOS when compared to Pt substitution. On the other hand, for $x=0.2$, the sharp jump emerges at $T_{c}\sim$ 2.45 K, indicating bulk superconductivity (Fig. 5(b)). According to the McMillan formula for electron-phonon mediated
superconductivity,\cite{McMillan} the electron-phonon coupling constant $\lambda $ can be determined by

\begin{equation}
T_{c}=\frac{\Theta _{D}}{1.45}\exp [-\frac{1.04(1+\lambda )}{\lambda -\mu
^{\ast }(1+0.62\lambda )}],
\end{equation}

where $\mu ^{\ast }\approx $ 0.13 is the common value for Coulomb
pseudopotential. By using $T_{c}$ = 2.45 K and $\Theta _{D}$ = 194.3 K, we obtain
$\lambda \approx$ 0.59, a typical value of weak-coupled BCS superconductor. The specific heat jump at $T_{c}$, $\Delta $C$_{es}$/$\gamma
T_{c}\approx$ 1.18, is somewhat smaller than the weak coupling value 1.43.\cite{McMillan} These results indicate that Ir$_{0.8}$Rh$_{0.2}$Te$_{2}$ is a weak-coupled BCS superconductor.

\begin{figure}[tbp]
\centerline{\includegraphics[scale=0.7]{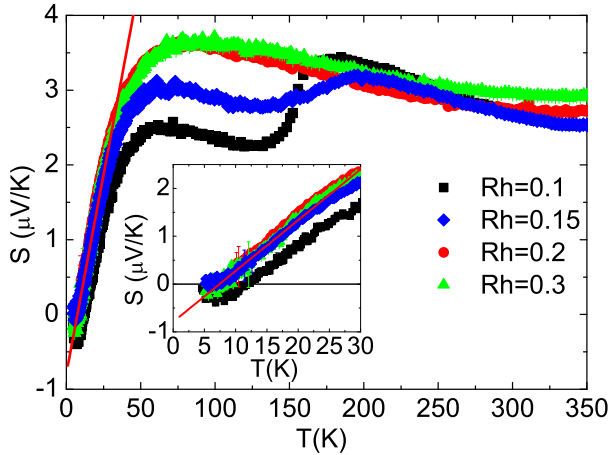}} \vspace*{-0.3cm}
\caption{ Temperature dependence of the Seebeck coefficient for Ir$_{1-x}$Rh$_{x}$Te$_{2}$. The red line is the linear fitting result as described in the text. Inset shows the low temperature region for clarity.}
\end{figure}

Finally, we show the Seebeck coefficient $S$ of Ir$_{1-x}$Rh$_{x}$Te$_{2}$ for $x=0.1-0.3$ in Fig. 6. Above about 15 K, the sign of $S$ for all samples is positive, indicating the hole-type carriers. The Seebeck coefficient of sample with $x=0.1$ shows anomaly at $\sim 170$ K which is consistent with the kink in resistivity in Fig. 3. In a metal with dominant single band transport and with diffusion mechanism and electron-type carriers, Seebeck coefficient is given by the Mott relationship,

\begin{eqnarray}
S=-\frac{\pi^2k_B^2T}{3e}\frac{\partial \ln\sigma(\mu)}{\partial \mu},
\end{eqnarray}

where $\rho(\varepsilon)$ is the DOS, $\varepsilon_F$ is the Fermi energy, $k_B$ is the Boltzman constant and $e$ is the absolute value of electronic charge.\cite{TE}  So the anomaly in $S$ should reflect the Fermi surface reconstruction related to the structural phase transition at same temperature. With increasing Rh concentration, the anomaly in Seebeck and resistivity is suppressed gradually and then disappears. This implies that the Rh substitution suppresses the structure phase transition and then the corresponding Fermi surface reconstruction. The low temperature parts of S(T) curve for Ir$_{0.9}$Rh$_{0.1}$Te$_{2}$ shows sign change below about 12 K. The temperature of sign change is suppressed below 5 K within our resolution for all superconducting samples ($x\geq0.15$) (Fig. 6(inset)). The S(T) curves for superconducting samples below about $\sim 30$ K and above 5 K are positive, linear and indistinguishable.\cite{Kudo} This implies that the Fermi surfaces do change somewhat as Rh enters the lattice, but only until superconductivity sets in. Together with the linear specific heat coefficient, this would imply that changes in the DOS at the Fermi level are visible only up to $x=$ 0.1, whereas further isovalent substitution of Rh on Ir site does not change the shape of the Fermi surface significantly.

\section{Conclusion}

In summary, we found that Rh substitution in Ir$_{1-x}$Rh$_{x}$Te$_{2}$ suppresses the first-order structural phase transition at high temperature. When $x>$ 0.15, the structural phase transition is absent and superconductivity sets in. Superconductivity is bulk and with the $T_{c}$ at $\sim$ 2.6 K by $x=$ 0.2. The substitution of Rh for Ir increases the electronic specific heat, i.e., likely recovers the lost area of the Fermi surface due to the reconstruction at high temperature phase transition. Unlike Pt or Pd substitution and Pd or Cu intercalation, the Rh substitution should not introduce extra carriers into Ir$_{1-x}$Rh$_{x}$Te$_{2}$. This testifies to competing relationship between structural transition and superconductivity and points to importance of structural parameters in high temperature phase transition suppression and emergence of bulk superconductivity.

\emph{Note added}. During the preparation of our manuscript we became aware that Kudo et al.\cite{Kudo} also reported superconductivity in Ir$_{1-x}$Rh$_{x}$Te$_{2}$ that crystallizes in CdI$_{2}$ - type structure. Their results are consistent with ours.

\section{Acknowledgements}

We thank the X7B at the NSLS Brookhaven Laboratory for the use of their equipments. Work at Brookhaven is supported by the U.S. DOE under Contract No. DE-AC02-98CH10886.

$^{\ast }$Present address: Frontier Research Center, Tokyo Institute of Technology,
4259 Nagatsuta, Midori, Yokohama 226-8503, Japan.


\begin{thebibliography}{99}

\bibitem{Wilson1} J. A. Wilson and A. D. Yoffe, Adv. Phys. \textbf{18}, 193 (1969).

\bibitem{Wilson2} J. A. Wilson, F. J. DiSalvo, and S. Mahajan, Adv. Phys. \textbf{24}, 117 (1975).

\bibitem{Castro} A. H. Castro Neto, Phys. Rev. Lett. \textbf{86}, 4382 (2001).

\bibitem{Valla} T. Valla, A. V. Fedorov, P. D. Johnson, J. Xue, K. E. Smith, F. J. DiSalvo, Phys. Rev. Lett. \textbf{85}, \textbf{4759} (2000).

\bibitem{Morosan}  E. Morosan, H. W. Zandbergen, B. S. Dennis, J. W. G. Bos, Y. Onose, T. Klimczuk, A. P. Ramirez, N. P. Ong, and R. J. Cava, Nature Phys. \textbf{2}, 544 (2006).

\bibitem{Liu} Y. Liu, R. Ang, W. J. Lu, W. H. Song, L. J. Li, and Y. P. Sun, Appli. Phys. Lett \textbf{102}, 192602 (2013).

\bibitem{Sipos} B. Sipos, A. F. Kusmartseva, A. Akrap, H. Berger, L. Forr\'{o}, and E. Tuti\v{s}, Nat. Mater. \textbf{7}, 960 (2008).

\bibitem{Gabovich} A. M. Gabovich, A. I. Voitenko, J. F. Annett, and M. Ausloos, Supercond. Sci. Technol. \textbf{14} R1 (2001).

\bibitem{Daou} R. Daou, J. Chang, David LeBoeuf, Olivier Cyr-Choiniere, Francis Laliberte, Nicolas Doiron-Leyraud, B. J. Ramshaw, Ruixing Liang, D. A. Bonn, W. N. Hardy and Louis Taillefer,Nature \textbf{463}, 519 (2010)

\bibitem{Tacon} M. Le Tacon, G. Ghiringhelli, J. Chaloupka, M. Moretti Sala, V. Hinkov,, M.W. Haverkort, M. Minola, M. Bakr, K. J. Zhou, S. Blanco-Canosa, C. Monney, Y. T. Song, G. L. Sun, C. T. Lin,
G. M. De Luca, M. Salluzzo, G. Khaliullin, T. Schmitt, L. Braicovich and B. Keimer, Nature Phys. \textbf{7}, 725 (2011)

\bibitem{Ghiringhelli} G. Ghiringhelli, M. Le Tacon, M. Minola, S. Blanco-Canosa, C. Mazzoli, N. B. Brookes, G. M. De Luca, A. Frano, D. G. Hawthorn, F. He, T. Loew, M. Moretti Sala, D. C. Peets, M. Salluzzo, E. Schierle, R. Sutarto, G. A. Sawatzky, E. Weschke, B. Keimer and L. Braicovich, Science. \textbf{337}, 821 (2012)

\bibitem{Torchinsky} D. H. Torchinsky, F. Mahmood, A. T. Bollinger, I. Bozovic and N. Gedik, Nature Mater. \textbf{12}, 387 (2013).

\bibitem{Chang} J. Chang, E. Blackburn, A. T. Holmes, N. B. Christensen, J. Larsen, J. Mesot, R. Liang, W. N. Hardy, A. Watenphul, M. v.Zimmermann, E. M. Forgan and S. M. Hayden, Nature Phys. \textbf{8} 871 (2012).

\bibitem{Pyon} S. Pyon, K. Kudo, and M. Nohara, J. Phys. Soc. Jpn. \textbf{81}, 053701 (2012).

\bibitem{Yang JJ} J. J. Yang, Y. J. Choi, Y. S. Oh, A. Hogan, Y. Horibe, K. Kim, B. I. Min, and S-W. Cheong, Phys. Rev. Lett. \textbf{108}, 116402 (2012).

\bibitem{Kamitani} M. Kamitani, M. S. Bahramy, R. Arita, S. Seki, T. Arima, Y. Tokura, and S. Ishiwata, Phys. Rev. B \textbf{87}, 180501(R) (2013).

\bibitem{Matsumoto} N. Matsumoto, K. Taniguchi, R. Endoh, H. Takano, and S. Nagata, J. Low Temp. Phys. \textbf{117}, 1129 (1999).

\bibitem{Ootsuki2} D. Ootsuki, Y. Wakisaka, S. Pyon, K. Kudo, M. Nohara, M. Arita, H. Anzai, H. Namatame, M. Taniguchi, N. L. Saini and T. Mizokawa, Phys. Rev. B. \textbf{86}, 014519 (2012).

\bibitem{Fang AF} A. F. Fang, G. Xu, T. Dong, P. Zheng, and N. L. Wang, Sci. Rep. \textbf{3}, 1153 (2013).

\bibitem{Mizuno} K. Mizuno, K.-i. Magishi, Y. Shinonome, T. Saito, K. Koyama, N. Matsumoto, S. Nagata, Physica B \textbf{312-313}, 818 (2002).

\bibitem{Ootsuki} D. Ootsuki, S. Pyon, K. Kudo, M. Nohara, M. Horio, T. Yoshida, A. Fujimori, M. Arita, H. Anzai, H. Namatame, M. Taniguchi, N. L. Saini, and T. Mizokawa, arXiv:1207.2613 (2012).

\bibitem{Kiswandhi} A. Kiswandhi, J. S. Brooks, H. B. Cao, J. Q. Yan, D. Mandrus, Z. Jiang, and H. D. Zhou, Phys. Rev. B \textbf{87}, 121107 (2013).

\bibitem{Oh} Y. S. Oh, J. J. Yang, Y. Horibe, and S.-W. Cheong, Phys. Rev. Lett. \textbf{110}, 127209 (2013).

\bibitem{Hammersley} A. P. Hammersley, S. O. Svenson, M. Hanfland, and D. Hauserman, High Pressure Res. \textbf{14}, 235 (1996).

\bibitem{Larson} A. C. Larson and R. B. V. Dreele, GSAS (General Structure Analysis System), Los Alamos National Laboratory Report No. LAUR 86-748, 2000 (unpublished).

\bibitem{Toby} B. H. Toby, J. Appl. Crystallogr. \textbf{34}, 210 (2001).

\bibitem{Young} R. A. Young, The Rietveld Method (Oxford University Press, Oxford, 1995).

\bibitem{Hockings} E. F. Hockings and J. G. White, J. Phys. Chem. \textbf{64}, 1042 (1960).

\bibitem{Lee1} C. S. Lee and G. J. Miller, Inorg. Chem. \textbf{38}, 5139 (1999).

\bibitem{Pettenkofer} C. Pettenkofer and W. Jaegermann, Phys. Rev. B \textbf{50}, 8816 (1994).

\bibitem{McMillan}  W. L. McMillan, Phys. Rev. \textbf{167}, 331 (1968).

\bibitem{TE} R. D. Barnard, \textit{Thermoelectricity in Metas and Alloys} (Taylor \& Francis, London, 1972).

\bibitem{Kudo} K. Kudo, M. Kobayashi, S. Pyon, and M. Nohara, arXiv:1307.4152 (2013).


\end{thebibliography}
\end{document}